\documentclass[aps,preprint,groupaddress]{revtex4}
\usepackage[dvipdfm,dvips]{graphicx}
\usepackage{amsmath}
\begin{document}
%
\title{\textbf{\large{Theory of the Conductivity in Semiconducting Single-Wall Carbon Nanotubes}}}

\author{S. Fujita, S. Moon, Y. Takato and James McNabb, III}
\affiliation{
	\small{Department of Physics, University at Buffalo, State University of New York,
		\centerline{Amherst, 14260, USA}
		\centerline{fujita@buffalo.edu}
	}
}
\author{S. Godoy}
\affiliation{
	\small{Departamento de F\'{i}sica, Facultad de Ciencias, Universidad Nacional Aut\'{o}noma de M\'{e}xico,
		\centerline{M\'{e}xico 04510, D.F., M\'{e}xico}
	}
}

\date{\today}

\begin{abstract}
The conduction of a single-wall carbon nanotube depends on the pitch. If there are an integral number of carbon hexagons per pitch, then the system is periodic along the tube axis and allows ``holes" (, and not ``electrons",) to move inside the tube. This case accounts for a semiconducting behavior with the activation energy of the order of around 3 meV. There is a distribution of the activation energy since the pitch and the circumference can vary. Otherwise nanotubes show metallic behaviors. ``Electrons" and ``holes" can move in the graphene wall (two dimensions). The conduction in the wall is the same as in graphene if the finiteness of the circumference is disregarded.
Cooper pairs formed by the phonon exchange attraction moving in the wall is shown to generate a temperature-independent conduction at low temperature (3 \text{--} 20 K).

\end{abstract}
\maketitle

\section{INTRODUCTION}
Iijima [1] found after his electron diffraction analysis that carbon nanotubes ranging 4 to 30 nanometers (nm) in diameter have helical multi-walled structures. Single-wall nanotube (SWNT) have about one nanometer in diameter and micrometers ($\mu$m) in length. Ebbesen et al. [2] measured the electrical conductivity $\sigma$ of carbon nanotubes and found that $\sigma$ varies depending of the temperature $T$, the tube radius $r$ and the pitch $p$. Experiments show that SWNT can be either semiconducting or metallic, depending on how they are rolled up from the graphene sheets [3]. In the present work we shall present a microscopic theory of the electrical conductivity of semiconducting SWNT, starting with a graphene honeycomb lattice, and developing a Bloch electron dynamics based on a rectangular cell model.

A SWNT can be formed by rolling a graphene sheet into a cylinder. The graphene which forms a honeycomb lattice is intrinsically anisotropic as we shall explain it in more detail later in Section II. Moriyama et al. [4] fabricated 12 SWNT devices from one chip, and observed that two of the SWNT samples are semiconducting and the other 10 are metallic, the difference in the room temperature resistance being of two to three orders of magnitude. The semiconducting SWNT samples show an activated-state temperature behavior. Why are there two sets of samples showing very different behavior? We believe the answer to this question arises as follows.

\begin{figure}[h] 
\includegraphics[scale=1]{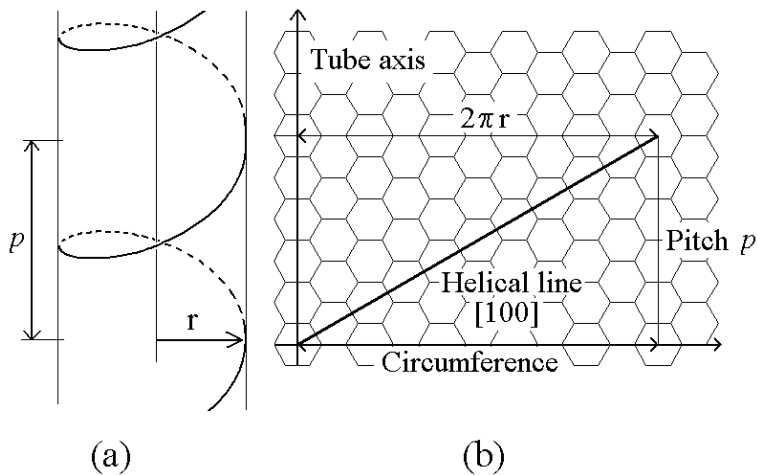}
\caption{(a) A section circular tube wall with a radius $r$ and a pith $p$. \\ (b) Its unrolled plane.}\label{f_honeycomb}
\end{figure}

The line passing the centers of the nearest-neighbor carbon hexagons forms a helical line around the nanotube with a pitch $p$. In Fig. 1 (a), a section of the circular tube with a pitch $p$ is drawn. Its unrolled plane is shown in (b). The circumference $2 \pi r$, where $r$ is the tube radius, likely contains an integral number of the carbon hexagons (units). The pitch $p$, however, may or may not contain an integral number of units. In the first alternative, the nanotube is periodic with the period $p$ along the tube axis. Then, there is a one-dimensional (1D) $k$-vector along the tube. A ``hole" which has a positive charge $+e$ and a size of a unit ring of height $p$ and radius $r$ can go through inside the positively charged carbon wall. An ``electron" having a negative charge $-e$ and a similar size is attracted by the carbon wall, and hence it cannot go straight inside the wall. Thus, there should be an extra ``hole" channel current in a SWNT. Moriyama et al. [4] observed a ``hole"-like current after examining the gate voltage effect. The system should have the lowest energy if the unit ring contains an integer set $(m,n)$ of carbon hexagons, which may be attained after annealing at high temperatures. This should happen if the tube length is comparable with the circumference. The experimental tube length is much greater than the circumference, and the pitch angle can be varied continuously. In the fabrication process the pitch is not controlled. The set of irrational numbers is greater in cardinality than the set of rational numbers. Then, the first case in which the unit contains an integer set $(m,n)$ of hexagons must be the minority. This case then generates a semiconducting transport behavior. We shall show later that the transport requires an activation energy. Fujita and Suzuki [5] showed that the ``electrons" and ``holes" must be activated based on the rectangular cell model for graphene.

If a SWNT contains an irrational number of carbon hexagons, which happens more often, then the system does not allow a conduction along the tube axis. The system is still conductive since the conduction electrons (``electrons", ``holes") can go through in the tube wall. This conduction is two-dimensional (2D) as can be seen in the unrolled configuration, which is precisely the graphene honeycomb lattice. This means that the conduction in the carbon wall should be the same as the conduction in graphene if the effect of the finiteness of the radius is neglected. 

We consider graphene in section II. The current band theory of the honeycomb crystal is based on the Wigner-Seitz (WS) cell model [3,6]. The WS model [6] is suitable for the study of the ground-state energy of the crystal. To describe the Bloch electron motion in terms of the mass tensor [7] a new theory based on the Cartesian unit cell not matching with the natural triangular crystal axes is more appropriate. The conduction electron moves as a wave packet formed by the Bloch waves as pointed out by Ashcroft and Mermin in their book [7]. This picture is fully incorporated in our new theoretical model. 

\section{graphene}

We consider a graphene which forms a 2D honeycomb lattice. The normal carriers in the electrical transport are ``electrons" and ``holes". The ``electron" (``hole") is a quasi-electron which has an energy higher than the Fermi energy \textit{and} which circulates clockwise (counterclockwise) viewed from the tip of the applied magnetic field vector. ``Electrons" (``holes") are excited on the positive (negative) side of the Fermi surface with the convention that the positive normal vector at the surface points in the energy-increasing direction.

We assume that the ``electron" (``hole") wave packet, called the ``electron" (``hole") hereafter, has the charge $-e \ (+e)$ and a size of a unit carbon hexagon, generated above (below) the Fermi energy $\epsilon_F$. We shall show that (a) the ``electron" and ``hole" have different internal charge distributions and different effective masses, (b) the ``electron" and ``hole" are thermally activated with different energy gaps $(\epsilon_1, \epsilon_2)$, (c) the ``electrons" and ``holes" move in different easy channels in which they travel, and (d) graphene is intrinsically anisotropic but the charge transport is isotropic if all channels are included into consideration.

\begin{figure}[h] 
\includegraphics[scale=0.8]{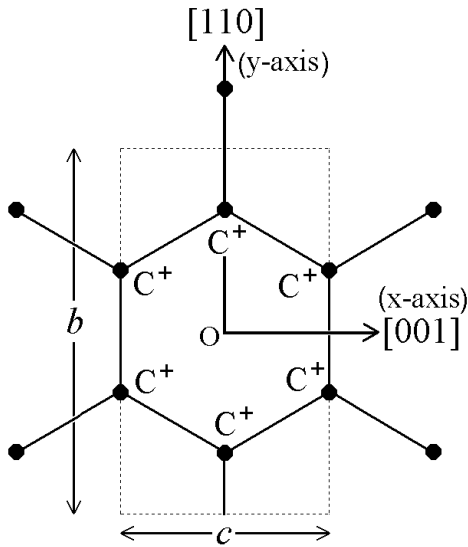}
\caption{A unit rectangular cell of graphene (dotted line).}\label{f_cell}
\end{figure}

The positively-charged ``hole" tends to stay away from positive ions C$^+$, and hence its charge is concentrated at the center of the hexagon. The negatively-charged ``electron" tends to stay close to the C$^+$ hexagon and its charge is concentrated near the C$^+$ hexagon. In our model, the ``electron" and ``hole" both have charge distributions, and they are not point particles. Hence, their masses $m_1$ and $m_2$ must be different from the gravitational mass $m = 9.11 \times 10^{-31}$ kg. Because of the different internal charge distributions, the ``electrons" and ``holes" should have the different effective masses $m_1$ and $m_2$. The ``electron" may easily move with a smaller effective mass in the direction [111 $c$-axis] $\equiv$ [110] than perpendicular to it as we see presently. Here, we use the conventional Miller indices for the hexagonal lattice with omission of the $c$-axis index [7]. For the description of the electron motion in terms of the mass tensor [7], it is convenient to introduce Cartesian coordinates, which do not match with the crystal’s natural (triangular) axes. We may choose the rectangular unit cell with side length $(b,c)$ as shown in Fig. 2. Note that graphene has a mirror (reflection) symmetry with respect to the $x$- and $y$-axis. The Brillouin zone in the $k$-space is unique: a rectangle with side lengths $(2 \pi /b, 2 \pi / c)$. The ``electron" (wave packet) may move up or down in [110] to the neighboring hexagon sites passing over one C$^+$. The positively charged C$^+$ acts as a welcoming (favorable) potential valley for the negatively charged ``electron". The same C$^+$ acts as a hindering potential hill for the positively charged ``hole". The ``hole" can however move easily sideways along the $x$-axis over on a series of vacant sites, each surrounded by six C, without meeting the hindering potential hills. Thus, the easy channel directions for ``electrons" and ``holes" are [110] and [001], respectively.

Let us consider the system (graphene) at 0 K. If we put an electron in the crystal, then the electron should occupy the center O of Brillouin zone, where the lowest energy lies. Additional electrons occupy points neighboring O in consideration of Pauli's exclusion principle. The electron distribution is lattice-periodic over the entire crystal in accordance with the Bloch theorem [8]. The uppermost partially filled bands are important for transport properties discussion. We consider such a band. The 2D Fermi surface which defines the boundary between the filled and unfilled $k$-space (area) is \textit{not} a circle since $x-y$ symmetry is broken $(b \neq c)$. The ``electron" effective mass is less in the direction [110] than perpendicular to it. That is, the system has two different masses and it is \textit{intrinsically anisotropic}. If the electron number is raised by the gate voltage applied perpendicular to the plane, then the Fermi surface more quickly grows in the easy-axis ($y$-) direction than in the $x$-direction. The Fermi surface must approach the Brillouin boundary at right angles because of the inversion symmetry possessed by honeycomb lattice [9]. Then at a certain voltage, a ``neck" Fermi surface must be developed. 

The same easy channels in which the ``electron" runs with a small mass, may be assumed for other hexagonal directions, [011] and [101]. The currents run in the three channels $\left< 110 \right> \equiv [110], [011]$, and $[101]$. The total current (magnitude) along the field direction $\mu$ is proportional to [10] 
\begin{align}
\sum_{\kappa \textrm{ channels}} \cos^2 (\mu, \kappa) = \cos^2 \theta + \cos^2 (\theta + 2 \pi / 3) + \cos^2 (\theta - 2 \pi / 3) = 3/2
\end{align}
Note that this current does not depend on the angle $\theta$ between the field direction ($\mu$) and the channel current direction ($\kappa$). More detailed discussion can be found in ref. [9]. Hence, the graphene does not show anisotropy in the conductivity.

We have seen earlier that the ``electron" and ``hole" have different internal charge distributions and therefore have different effective masses. Which carriers are easier to be activated or excited? This question can be answered without considering  channeling. The ``electron" is near the positive ions and the ``hole" is farther away from the ions. Hence, the gain in the Coulomb interaction is greater for the ``electron". That is, the ``electrons" are more easily activated (or excited). The presence of the welcoming C$^+$ ions in the channel direction also enhances this inequality. We may represent the activation energy difference by 
\begin{align}
\epsilon_1 < \epsilon_2.
\end{align}
The thermally-activated (or excited) electron densities are given by 
\begin{align}
n_{j} (T) = n_{j} e^{\epsilon_{j} / k_B T},
\end{align}
where $j = 1$ and 2 represent the ``electron" and ``hole", respectively. The pre-exponential factor $n_{j}$ is the density at high-temperature limit.

\section{SINGLE-WALL NANOTUBES}

Let us consider the long SWNT rolled with a graphene sheet. The charge may be transported by the channeling ``electrons" and ``holes" in the graphene wall. But the ``holes" within the wall can also contribute to the charge transport. Because of this extra channel inside the carbon nanotube, ``holes" can be the majority carriers in nanotubes although ``electrons" are the dominant carriers in graphene. Moriyama et al. [4] studied the electrical transport in semiconducting SWNT in the temperature range 2.6 - 200 K, and found from the filed (gate voltage) effect that the carriers are ``hole"-like. Their data are reproduced in Fig. 3.

\begin{figure}[h] 
\includegraphics[scale=1.3]{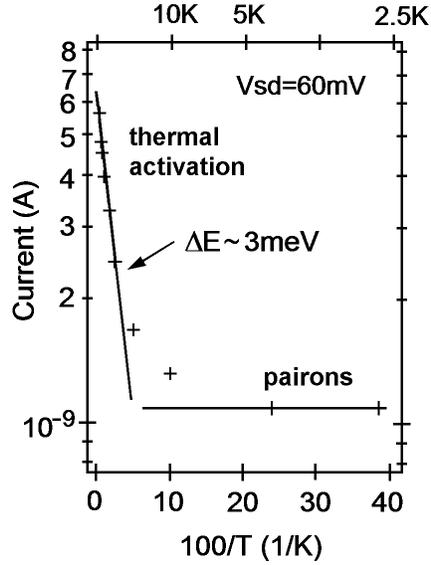}
\caption{Log-scale plot of the currents in semiconducting SWNT as a function of inverse temperature after Moriyama et al. [4].}\label{f_moriyama}
\end{figure}

The conductivity depends on the pitch of the SWNT. The helical line is defined as the line passing the nearest neighbors of the C-hexagons. The helical angle $\theta$ is the angle between the helical line and the tube axis. The degree of helicity $h$ may be defined as 
\begin{align}
h \equiv \cos \theta.
\end{align}
For a macroscopically large graphene the conductivity is isotropic as we saw in Section II. The conductivity $\sigma$ in (semiconducting) SWNT depends on this helicity $h$. This is a kind of a finite size effect.. The circumference is finite while the tube length is macroscopic. 

In a four-valance-electron conductor such as graphene all electrons are bound to ions, and there is no conduction at 0 K. If a ``hole" having the charge $+e$ and the size of a unit ring is excited, then this ``hole" can move along the tube axis with the activation energy $\epsilon_3$ and the effective mass $m_3$. Both $\epsilon_3$ and $m_3$ depend on the radius and the pitch.

We are now ready to discuss the conductivity of semiconducting SWNT. There are four currents carried by 
\\ (a)	``Electrons" moving in the graphene wall with the mass $m_1$ and the density $n_1 \exp (- \epsilon_1 / k_B T)$, running in the channels $\left< 110 \right>$.
\\ (b)	``Holes" moving in the graphene wall with the mass $m_2$ and the density $n_2 \exp (- \ \epsilon_2 / k_B T)$, running in the channels $\left< 100 \right>$.
\\ (c)	``Holes" moving with the mass $m_3$ and the density $n_3 \exp (- \epsilon_3 / k_B T)$, running in the tube-axis direction. The activation  (or excitation) energy $\epsilon_3$ and the effective mass $m_3$ vary with the radius and the pitch. 
 \\(d)	Cooper pairs (pairons) formed by the phonon-exchange attraction, which move in the graphene wall.
 
In actuality, one of the currents may be dominant, and be observed.

In the normal Ohmic conduction due to the conduction electrons the resistance is proportional to the sample (tube) length. Then, the conductivity $\sigma$ is given by the Drude formula:
\begin{align}
\sigma = \frac{q^2}{m^*} n \tau \equiv \frac{q^2}{m^*} n \frac{1}{\Gamma},
\end{align}
where $q$ is the carrier charge $(\pm e), m^*$ the effective mass, $n$ the carrier density, and $\tau$ the relaxation (collision) time. The relaxation rate $\Gamma \equiv \tau^{-1}$ is the inverse of the relaxation time. If the impurities and phonons are scatterers, then the rate $\Gamma$ is the sum of the impurity scattering rate $\Gamma_{\textrm{imp}}$ and the phonon scattering rate $\Gamma_{\textrm{ph}}(T)$:
\begin{align}
\Gamma = \Gamma_{\textrm{imp}} + \Gamma_{\textrm{ph}}(T).
\end{align}
The impurity scattering rate $\Gamma_{\textrm{imp}}$ is temperature-independent and the phonon scattering rate $\Gamma_{\textrm{ph}}(T)$ is temperature $
(T)$-dependent. The phonon scattering rate $\Gamma_{\textrm{ph}}(T)$ is linear in $T$ above around 2 K:
\begin{align} 
\Gamma_{\textrm{ph}}(T) = aT, \quad a = \textrm{constant, (above 2 K)}.
\end{align}
The temperature dependence should arise from the carrier density $n(T)$ and the phonon scattering rate  $\Gamma_{\textrm{ph}}(T)$. Writing the $T$-dependence explicitly, we obtain from (3) and (7)
\begin{align}
\sigma = \sum_j \frac{e^2}{m^*_j} n_j e^{-\epsilon_j / k_B T} \frac{1}{\Gamma_{\textrm{imp}}+aT}.
\end{align}
Moriyama et al. [4] used the Arrhenius plot for the data above 20 K and obtained the activation energy
\begin{align}
\epsilon_3 \sim 3 \textrm{ meV}.
\end{align}
By studying the field (gate voltage) effect, the carriers were found to be ``hole"-like. Thus, the major currents observed can be interpreted in terms of the ``holes" moving within the tube wall.

This ``hole" axial transport depends on the unit ring containing $m \times n$ hexagons. Since the pitch and the circumference have distributions, the activation energy $\epsilon_3$ should also have a distribution. Hence the obtained value in Eq. (9) must be regarded as the averaged value.

Liu et al. [10] systematically measured the resistance $\rho(T)$ of SWNT under hydrostatic pressures, and fitted their data by using 2D \textit{variable range hopping} (vrh) theoretical formula [11]:
\begin{align}
\rho (T) &= \rho_0 \exp (T_0/T)^{1/3},
\end{align}
where 
\begin{align}
\quad T_0 &= 525 \ \mathrm{K}.
\end{align}
is a fit (temperature) parameter and $\rho_0$ a (resistance) parameter. Mott's vrh theory [11] is applicable when highly random disorders exist in the system. An individual SWNT (annealed) is unlikely to have such randomness. We take a different view here. The scatterings are due to normally assumed impurities and phonons. But carriers (``holes") have a distribution in the unit cell size. Hence the distribution of the activation energy introduces the flattening of the Arrhenius slope by the factor 1/3. Compare Eqs. (10) and (3).

The ``hole" size is much greater than the usually assumed atomic impurity size and the phonon size, which are of the order of the lattice constant. This size mismatch may account for a ballistic charge transport observed by Frank et al. [13] and others [14]. More careful studies are required to establish the cause of the ballisticity.

We now go back to the data shown in Fig. 3. Below 20 K the currents observed are very small and they appear to approach a constant in the low temperature limit (large $T^{-1}$ limit). These currents, we believe, are due to the Cooper pairs.

The Cooper pairs (pairons) move in 2D with the linear dispersion relation [15]:
\begin{align}
\epsilon &= c^{(j)} p \\
c^{(j)} &= \frac{2}{\pi} v_F^{(j)},
\end{align}
where $v_F^{(j)}$ is the Fermi velocity of the ``electron" ($j = 1
$)  [``hole" ($j = 2$)]. 

Consider first ``electron"-pairs. The velocity $\mathbf{v}$ is given by (omitting superscript)
\begin{align}
\mathbf{v} &= \frac{\partial \epsilon}{\partial \mathbf{p}} \quad \textrm{or}
 \quad v_x = \frac{\partial \epsilon}{\partial p} \frac{\partial p}{\partial p_x} 
= c \frac{p_x}{p} \\
p &\equiv \sqrt{p_x^2 + p_y^2}.
\end{align}
The equation of motion along the $E$-field ($x$-) direction is
\begin{align}
\frac{\partial p_x}{\partial t} = q' E,
\end{align}
where $q'$ is the charge $\pm 2e$ of a pairon. The solution of Eq. (16) is given by 
\begin{align}
p_x = q'Et + p_x^{(0)},
\end{align}
where $ p_x^{(0)}$ is the initial momentum component. The current density $j_p$ is calculated from (charge $q'$)  $\times$ (number density $n_p$) $\times$ (average velocity $\bar{v}$). The average velocity $\bar{v}$ is calculated by using Eq. (14) and Eq. (17) with the assumption that the pair is accelerated only for the collision time $\tau$. We obtain
\begin{align}
j_p &\equiv q' n_p \bar{v}
= q' n_p c \frac{1}{p} (q' E \tau) = q'^2 \frac{c}{p} n_p E \tau.
\end{align}

For stationary currents, the pairon density $n_p$ is given by the Bose distribution function $f(\epsilon_p)$
\begin{align}
n_p = f(\epsilon_p) \equiv [\exp(\epsilon_p / k_B T - \alpha) - 1]^{-1},
\end{align}
where $e^{\alpha}$ is the fugacity.
Integrating the current $j_p$ over all 2D $p$-space, and using Ohm’s law $j = \sigma E$ we obtain for the conductivity $\sigma$:
\begin{align}
\sigma = (2 \pi \hbar)^{-2} q'^2 c \int d^2p \ p^{-1} f(\epsilon_p) \tau.
\end{align}
In the temperature ranging between 2 and 20 K we may assume the Boltzmann distribution function for $f(\epsilon_p)$:
\begin{align}
f(\epsilon_p) \simeq \exp{(\alpha - \epsilon_p / k_B T)},
\end{align}
We assume that the relaxation time arises from the phonon scattering so that $\tau = (aT)^{-1}$, see Eqs. (5)-(7). After performing the $p$-integration we obtain
\begin{align}
\sigma = \frac{2}{\pi} \frac{e^2 k_B}{a \hbar^2} e^\alpha,
\end{align}
which is temperature-independent. If there are ``electrons" and ``hole" pairs, they contribute additively to the conductivity.
These pairons should undergo a Bose-Einstein condensation at a temperature lower than 2.2 K. We predict a superconducting state at lower temperatures.

\section{Summary and Discussion}

A SWNT is likely to have an integral number of carbon hexagons around the circumference. If each pitch contains an integral number of hexagons, then the system is periodic along the tube axis, and ``holes" (and not ``electrons") can move along the tube axis. The system is semiconducting with an activation energy $\epsilon_3$. This energy $\epsilon_3$ has a distribution since both pitch and circumference have distributions. The pitch angle is not controlled in the fabrication processes. There are more numerous cases where the pitch contains an irrational numbers of hexagons. In these cases the system shows a metallic behavior experimentally [15]. 

In the process of arriving at our main conclusion we have uncovered the following results.
\\ (a)	``Electrons" and ``holes" can move in 2D in the carbon wall in the same manner as in graphene.
\\ (b)	For a metallic SWNT (a) implies that the conduction in the wall shows no pitch dependence.
\\ (c)	The Cooper pairs are formed in the wall. They should undergo Bose-Einstein condensation at lower temperature, exhibiting a superconducting state.

A metallic SWNT will be treated separately.\\

\textbf{REFERENCE} \\
$[1]$ S. Iijima, Nature (London) \textbf{354}, 56 (1991). \\
$[2]$ T. W. Ebbesen, H. J. Lezec, H. Hiura, J. W. Bennet, H. F. Ghaemi and T. Thio, Nature \textbf{382}, 54 (1996). \\
$[3]$ R. Saito, G. Dresselhaus and M. S. Dresselhaus, \textit{Physical Properties of Carbon Nanotubes} (Imperial College Press, London, 1998) pp. 35-58. \\
$[4]$ S. Moriyama, K. Toratani, D. Tsuya, M. Suzuki, Y. Aoyagi and K. Ishibashi, Physica E \textbf{24}, 46 (2004). \\
$[5]$ S. Fujita and A. Suzuki, J. Appl. Phys. \textbf{107}, 013711 (2010). \\
$[6]$ E. Wigner and F. Seitz, Phys. Rev. \textbf{43}, 804 (1933). \\
$[7]$ N. W. Ashcroft and N. D. Mermin, \textit{Solid State Physics} (Saunders, Philadelphia, 1976), p.217, pp. 91-93, pp. 228-229. \\
$[8]$ F. Bloch, Zeits, f. Physik, \textbf{52}, 555 (1928). \\
$[9]$ S. Fujita, and K. Ito, \textit{Quantum Theory of Conducting Matter} (Springer, New York, 2007), pp. 106-107, pp. 85-90. \\
$[10]$ S. Fujita, A. Garcia, D. O'Leyar, S. Watanabe and T. Burnett, J. Phys. Chem. Solids \textbf{50}, 27 (1989). \\
$[11]$ B. Liu, et al. , Solid State Commun. \textbf{118}, 31 (2001). \\
$[12]$ N. F. Mott, \textit{Conduction in Non-Crystalline Materials} (Oxford University Press, Oxford, 1987). \\
$[13]$ S. Frank, P. Poncharal, Z. I. Wang, and W. A. de Heer, Science \textbf{280}, 1744 (1998). \\
$[14]$ A. Javey, J. Guo, Q. Wang, M. Lundstrom, and H. Dai, Nature \textbf{424}, 654 (2003). \\
$[15]$ S. Fujita, K. Ito, and S. Godoy, \textit{Quantum Theory of Conducting Matter: Superconductivity} (Springer, New York, 2009), pp. 77-79. \\
$[16]$ S. J. Tans, M. H. Devoret, H. Dai, A. Thess, R. E. Smalley, L. J. Geerligs, and C. Dekker, Nature (London) \textbf{386}, 474 (1997).


\end{document}